\documentclass[aps,prl,preprint,nopacs,superscriptaddress]{revtex4}
\usepackage{sidecap}
\usepackage{graphicx}
\usepackage{verbatim}
\usepackage{mathrsfs}
\usepackage{amsmath,amsfonts,amssymb}
\usepackage{graphicx}

\def\3{2.8in}    %used for figure widths
\def\2{2.5in}
\def\4{3.0in}

\def \beq {\begin{equation}}
\def \eeq {\end{equation}}
\begin{document}

\title{Quasi-particle interferences of the Weyl semimetals TaAs and NbP}
\author{Guoqing Chang$^*$}\affiliation{Centre for Advanced 2D Materials and Graphene Research Centre National University of Singapore, 6 Science Drive 2, Singapore 117546}\affiliation{Department of Physics, National University of Singapore, 2 Science Drive 3, Singapore 117542}
\author{Su-Yang Xu$^*$}\affiliation {Laboratory for Topological Quantum Matter and Spectroscopy (B7), Department of Physics, Princeton University, Princeton, New Jersey 08544, USA}
\author{Hao Zheng\footnote{These authors contributed equally to this work.}}\affiliation {Laboratory for Topological Quantum Matter and Spectroscopy (B7), Department of Physics, Princeton University, Princeton, New Jersey 08544, USA}

\author{Chi-Cheng Lee}
\affiliation{Centre for Advanced 2D Materials and Graphene Research Centre National University of Singapore, 6 Science Drive 2, Singapore 117546}
\affiliation{Department of Physics, National University of Singapore, 2 Science Drive 3, Singapore 117542}
\author{Shin-Ming Huang}
\affiliation{Centre for Advanced 2D Materials and Graphene Research Centre National University of Singapore, 6 Science Drive 2, Singapore 117546}
\affiliation{Department of Physics, National University of Singapore, 2 Science Drive 3, Singapore 117542}

\author{Ilya Belopolski}\affiliation {Laboratory for Topological Quantum Matter and Spectroscopy (B7), Department of Physics, Princeton University, Princeton, New Jersey 08544, USA}
\author{Daniel S. Sanchez}\affiliation {Laboratory for Topological Quantum Matter and Spectroscopy (B7), Department of Physics, Princeton University, Princeton, New Jersey 08544, USA}

\author{Guang Bian}\affiliation{Laboratory for Topological Quantum Matter and Spectroscopy (B7), Department of Physics, Princeton University, Princeton, New Jersey 08544, USA}
\author{Nasser Alidoust}\affiliation {Laboratory for Topological Quantum Matter and Spectroscopy (B7), Department of Physics, Princeton University, Princeton, New Jersey 08544, USA}

\author{Tay-Rong Chang}\affiliation{Department of Physics, National Tsing Hua University, Hsinchu 30013, Taiwan}

\author{Chuang-Han Hsu}\affiliation{Centre for Advanced 2D Materials and Graphene Research Centre National University of Singapore, 6 Science Drive 2, Singapore 117546}
\affiliation{Department of Physics, National University of Singapore, 2 Science Drive 3, Singapore 117542}

\author{Horng-Tay Jeng}\affiliation{Department of Physics, National Tsing Hua University, Hsinchu 30013, Taiwan}\affiliation{Institute of Physics, Academia Sinica, Taipei 11529, Taiwan}

\author{Arun Bansil}\affiliation{Department of Physics, Northeastern University, Boston, Massachusetts 02115, USA}

\author{Hsin Lin$^{\dag}$}
\affiliation{Centre for Advanced 2D Materials and Graphene Research Centre National University of Singapore, 6 Science Drive 2, Singapore 117546}
\affiliation{Department of Physics, National University of Singapore, 2 Science Drive 3, Singapore 117542}

\author{M. Zahid Hasan\footnote{Corresponding authors: nilnish@gmail.com, mzhasan@princeton.edu}}\affiliation {Laboratory for Topological Quantum Matter and Spectroscopy (B7), Department of Physics, Princeton University, Princeton, New Jersey 08544, USA}

\begin{abstract}
The recent discovery of the first Weyl semimetal in TaAs provides the first observation of a Weyl fermion in nature. Such a topological semimetal features a novel type of anomalous surface state, the Fermi arc, which connects a pair of Weyl nodes through the boundary of the crystal. Here, we present theoretical calculations of the quasi-particle interference (QPI) patterns that arise from the surface states including the topological Fermi arcs in the Weyl semimetals TaAs and NbP. Most importantly, we discover that the QPI exhibits termination-points that are fingerprints of the Weyl nodes in the interference pattern. Our results, for the first time, propose an interference signature of the topological Fermi arcs in TaAs, which provides important guidelines for STM measurements on this prototypical Weyl semimetal compound. The scattering channels presented here is relevant to transport phenomena on the surface of the TaAs class of Weyl semimetals. Our work is also the first systematic calculation of the quantum interferences from the Fermi arc surface states, which is in general useful for future STM studies on other Weyl semimetals.

%On the other hand, due to the indirect nature of the QPI measurements, we show that the QPI pattern cannot prove for the Fermi arcs without reliance on calculations, in contrast to the case of photoemission experiments. 
\end{abstract}

\date{\today}
%In contrast to the spin texture of prototypical topological insulators where spin rotates with the surface contour in a tangential way, the spin of Fermi arcs in TaAs rotates against the contour, giving rise to a Dresselhaus spin texture. 
\maketitle
Weyl fermion semimetals are an exciting frontier of condensed matter physics and materials science. Such a crystal hosts Weyl fermions quasi-particles in the electronic band structure and admits a topological classification beyond band insulators \cite{rev2}. It has deep analogies with particle physics and leads to unique topological properties \cite{Weyl1, Weyl2, Wan, Weyl3, Weyl4, Ojanen, arc1, arc2}. Specifically, the Weyl fermions correspond to points of bulk band degeneracy, Weyl nodes. Each Weyl node has a definite chirality or chiral charge, which is a monopole or anti-monopole of Berry curvature in momentum space. The chiral charge is associated with an integer-valued topological index. This guarantees a new topological surface state, a Fermi arc, which connects the Weyl nodes through the boundary of the sample. In contrast to topological insulators, both the surface and the bulk of Weyl semimetals can give rise to new physics, opening up wide-ranging research opportunities. In the bulk, a Weyl semimetal crystal paves the way for studying the properties of the elusive Weyl fermion particle in high energy physics in table-top experiments. The presence of parallel electrical and magnetic fields can break the apparent conservation of the chiral charge due to the chiral anomaly, making a Weyl semimetal, unlike ordinary nonmagnetic metals, more conductive with an increasing magnetic field\cite{Weyl4, CA1, CA2}. Chiral photons can couple differently to Weyl fermions of opposite chiralities, leading to a spontaneous anomalous Hall current \cite{Photon}. The surface Fermi arcs feature a new type of quantum oscillation in transport, where electrons move in real space between different surfaces of a bulk sample when executing a constant-energy orbit in momentum space under an external magnetic field \cite{Ojanen, arc1, arc2}. These phenomena make new physics accessible and suggest potential applications.

The recent discovery of Weyl semimetal TaAs provided the first material realization of this new phase of matter \cite{TaAs1, TaAs2, ARPES1}. Both the Weyl fermions and the Fermi arcs have been directly observed in TaAs by photoemission experiments \cite{ARPES1}. Following the discovery, later ARPES results cemented the Weyl state in TaAs and studied the other three compound in the same family, namely NbAs, TaP and NbP \cite{ARPES2, ARPES3, ARPES4, TaAs_spin1, TaAs_spin2, NbAs, TaP1, TaP2, TaP3, NbP1, NbP2, NbP3}. On the other hand, scanning tunneling microscopy (STM) experiments had been lacking. Only very recently, the first STM study of the Weyl semimetal NbP has been reported \cite{Hao}. Historically, STM has been proven as a very powerful tool in the fields of high $T_{\textrm{c}}$ superconductors, graphene and topological insulators due to its simultaneous spatial, energy, and (quasi-)momentum resolution \cite{HTC, C, STM1, STM2, STM3}.

In this paper, we theoretically compute the quasi-particle interference patterns (QPIs) that arise from the surface states of the Weyl semimetals TaAs and NbP including the topological Fermi arcs. Our results answer following important questions: (1) What is the configuration of the QPI? (2) What are the scattering channels that lead to the observed dominant features in the QPI? (3) Is there any feature associated with the topological Fermi arcs? (4) Is there any feature associated with the Weyl nodes, i.e. the $k$ space locations where the Fermi arcs are terminated? In general, our results provide crucial theoretical information for any future STM studies on Weyl semimetals. Moreover, the scattering channels theoretically uncovered here has important implications for surface transport of Weyl semimetals.

\begin{figure*}
\centering
\includegraphics[width=16cm]{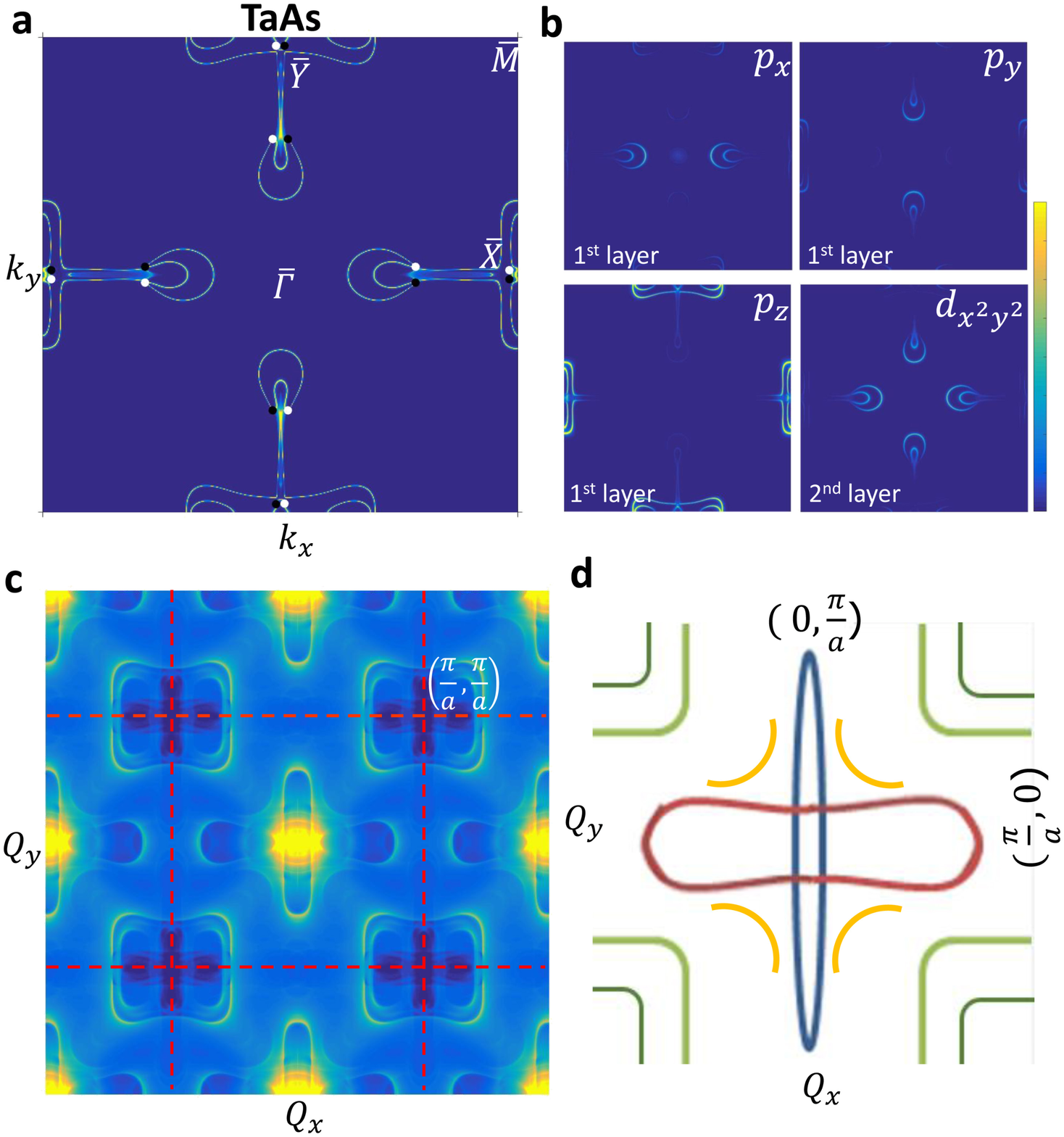}
\caption{\label{TaAsQPI}
\textbf{Theoretically calculated surface Fermi surface and QPI pattern on TaAs(001) surface.}
(a) Calculated (001) surface Fermi surface of TaAs.
%The high symmetry points are marked.
The black and white dots indicate the projected Weyl nodes with positive and negative chiral charges.
(b) Electronic states on the Fermi surface that arise from different orbitals. The first layer is As whereas the second layer is Ta.
(c) Calculated QPI pattern based on the Fermi surface in panel (a).
(d) A sketch of the QPI pattern that corresponds to the real calculation in panel (c).}
\end{figure*}

Figure~\ref{TaAsQPI}(a) shows the theoretical calculated As-terminated (001) surface Fermi surface of TaAs (001). We consider the pnictide termination throughout this work as it is the natural cleavage found in all experiments \cite{ARPES2, ARPES3, ARPES4, TaAs_spin1, TaAs_spin2, NbAs, TaP1, TaP2, TaP3, NbP1, NbP2, NbP3, Hao}. The calculated surface state Fermi surface is in excellent agreement with our ARPES data on TaAs \cite{ARPES1}. We identify three prominent features, namely, an elliptical feature at the $\bar{X}$ point, a bow-tie shaped contour at the $\bar{Y}$ points, and a crescent-shaped contour near the midpoint of the $\bar{\Gamma}-\bar{X}$ or $\bar{\Gamma}-\bar{Y}$ line. Due to the close proximity of the Weyl nodes near the $\bar{X}$ ($\bar{Y}$) point, the corresponding Fermi arc is extremely short and hence does not have any observable effects to the QPI pattern. On the other hand, the crescent feature consists of  Fermi arcs that join each other at the two end points, which correspond to projected Weyl nodes with projected chiral charge of $\pm2$. We further study the orbital characters of the crescent Fermi arcs. As shown in Fig.~\ref{TaAsQPI}(b), the crescent Fermi arcs arise from the $p_x$ and $p_y$ orbitals from the first layer, the As atoms and $d_{x^2y^2}$ orbital from the second later, the Ta atoms. Fig.~\ref{TaAsQPI}(c) shows the QPI pattern calculated from the surface band spectra in (a). It shows a rich structure, indicating that the scattering behavior on the TaAs surface is complicated. We sketch the dominant features in Fig.~\ref{TaAsQPI}(d). In the origin of the QPI image, we find an elliptical contour and a bowtie-shaped contour, whose long axes are perpendicular to each other. At each corner of the QPI, we observe two concentric squares. In addition, we also find weak features that seem to be open curves in each quadrant, as noted by the yellow curves.

\begin{figure*}
\centering
\includegraphics[width=16cm]{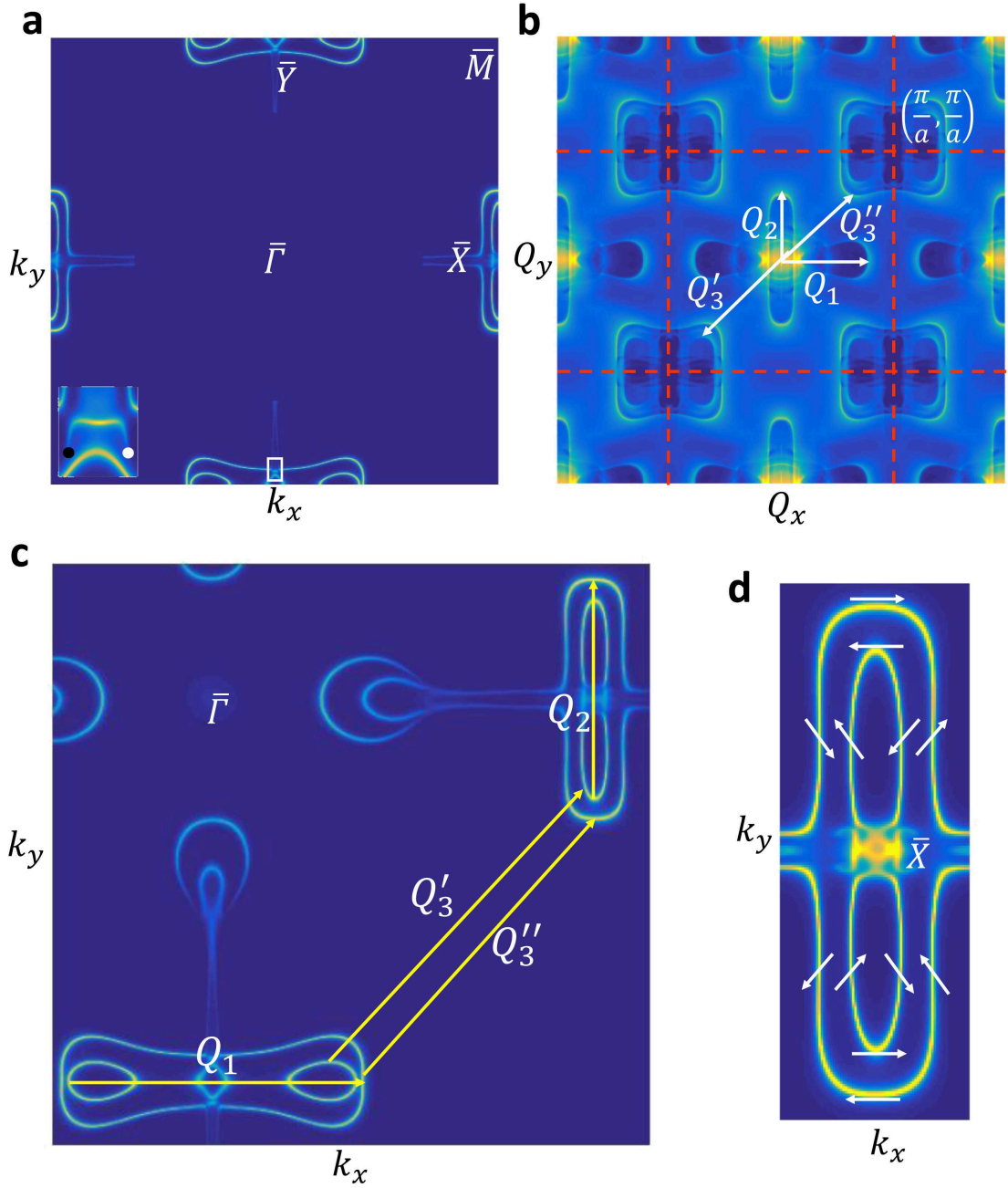}
\caption{\label{Trivial}
\textbf{Quasi-particle scattering that arises from trivial surface states.}
(a) Calculated surface Fermi surface where the crescent shaped Fermi arcs are manually removed. The inset shows a zoomed-in view of the $k$ space region highlighted by the white box, which encloses a pair of Weyl nodes near the surface BZ boundary $\bar{Y}$ point. It can be seen that the Fermi arc is a very short line that directly connects the pair of nodes. Since the pair of nodes are too close to each other, the Fermi arc does not have any significant impact to the calculated QPI pattern.
(b) Theoretical QPI pattern based on the Fermi surface in panel (a).
Four characteristic scattering vectors ($Q_1$, $Q_2$, $Q_3^{'}$, and $Q_3^{''}$) are shown.
(c) The four scattering vectors ($Q_1$, $Q_2$, $Q_3^{'}$, and $Q_3^{''}$) in $k-$space.
(d) A sketch of the spin texture of the elliptical surface Fermi contours at the $\bar{X}$ point. }
\end{figure*}

We study the scattering channel for the dominant features in the QPI. In Fig.~\ref{Trivial}, we only consider the bowtie-shaped and elliptical features at the surface Brillouin zone (BZ) boundaries by manually removing the crescent-shaped Fermi arcs from the Fermi surface. The Fermi arcs near the BZ boundaries are too short to have any real impact. The calculated QPI pattern based on this modified Fermi surface is shown in Fig.~\ref{Trivial}(b), where almost all dominate features in the full pattern in Fig.~\ref{TaAsQPI}(c) are reproduced except the weak features noted by the yellow curves in Fig.~\ref{TaAsQPI}(d). In order to understand the origin of these QPI features, we consider possible scattering channels of the Fermi surface. We consider the following scattering vectors, $Q_1$, $Q_2$, $Q_3^{'}$ and $Q_3^{''}$, as shown in Fig.~\ref{TaAsQPI}(b). By comparing the vector lengths in $k$-space and in $Q$-space, one can figure out the scattering channels. From Fig.~\ref{Trivial}(c), we clearly resolve that $Q_1$ and $Q_2$ are intra-contour-scattering within a bowtie-shaped or an elliptical feature in the Fermi surface, while $Q_3^{'}$ and $Q_3^{''}$ are the inter-contour-scattering between a bowtie-shaped feature and an elliptical feature. More importantly, the elliptical and bow-tie shaped features in the Fermi surface ($k$-space) and in the QPI pattern ($Q$-space) have almost identical line shapes. This similarity makes the identification of the QPI feature quite straightforward and reliable. In addition, one may notice that the elliptical feature in the Fermi surface ($k$-space) consists of two concentric contours at each $\bar{X}$ point but the resulting elliptical feature in the QPI ($Q$-space) is only one-fold. This is due to the fact that the elliptical feature in the Fermi surface is located at the $\bar{X}$ point, which is a time-reversal invariant Kramers' point. Hence the spin texture (Fig.~\ref{Trivial}(d)) requires that the scattering can only occur in-between the inner and the outer elliptical contour in the Fermi surface.

\begin{figure*}
\centering
\includegraphics[width=16cm]{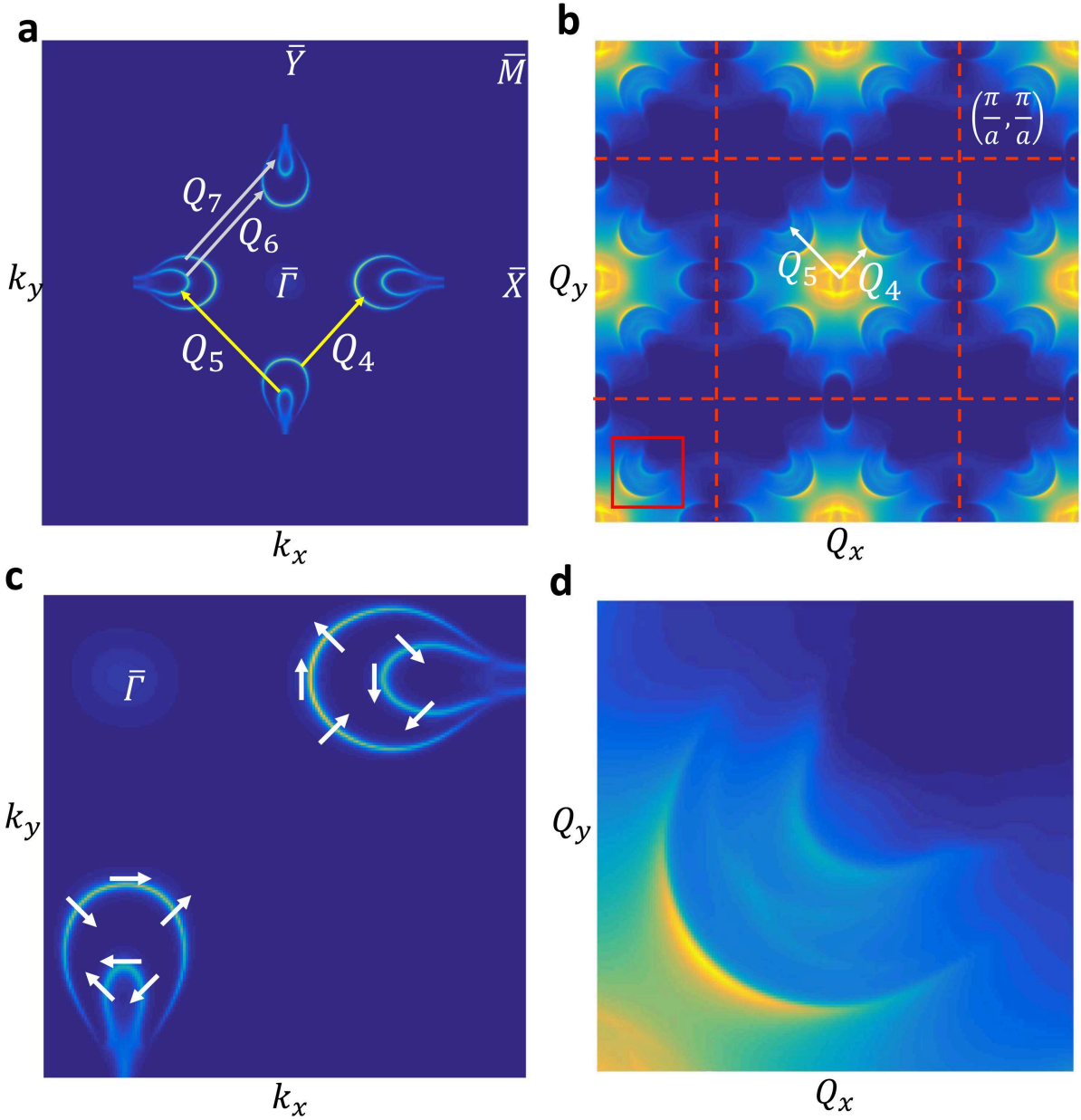}
\caption{\label{Nontrivial}
\textbf{Quasi-particle scattering that arises from the topological Fermi arcs.}
(a) Calculated surface Fermi surface containing only the crescent Fermi arcs.
(b) Theoretical QPI pattern based on panel (a).
(c) A sketch of the spin texture of the crescent Fermi arcs.
(d) A close-up view of the complex QPI feature that arise from the crescent Fermi arcs in one quadrant in $Q-$space.}
\end{figure*}

\begin{figure*}
\centering
\includegraphics[width=16cm]{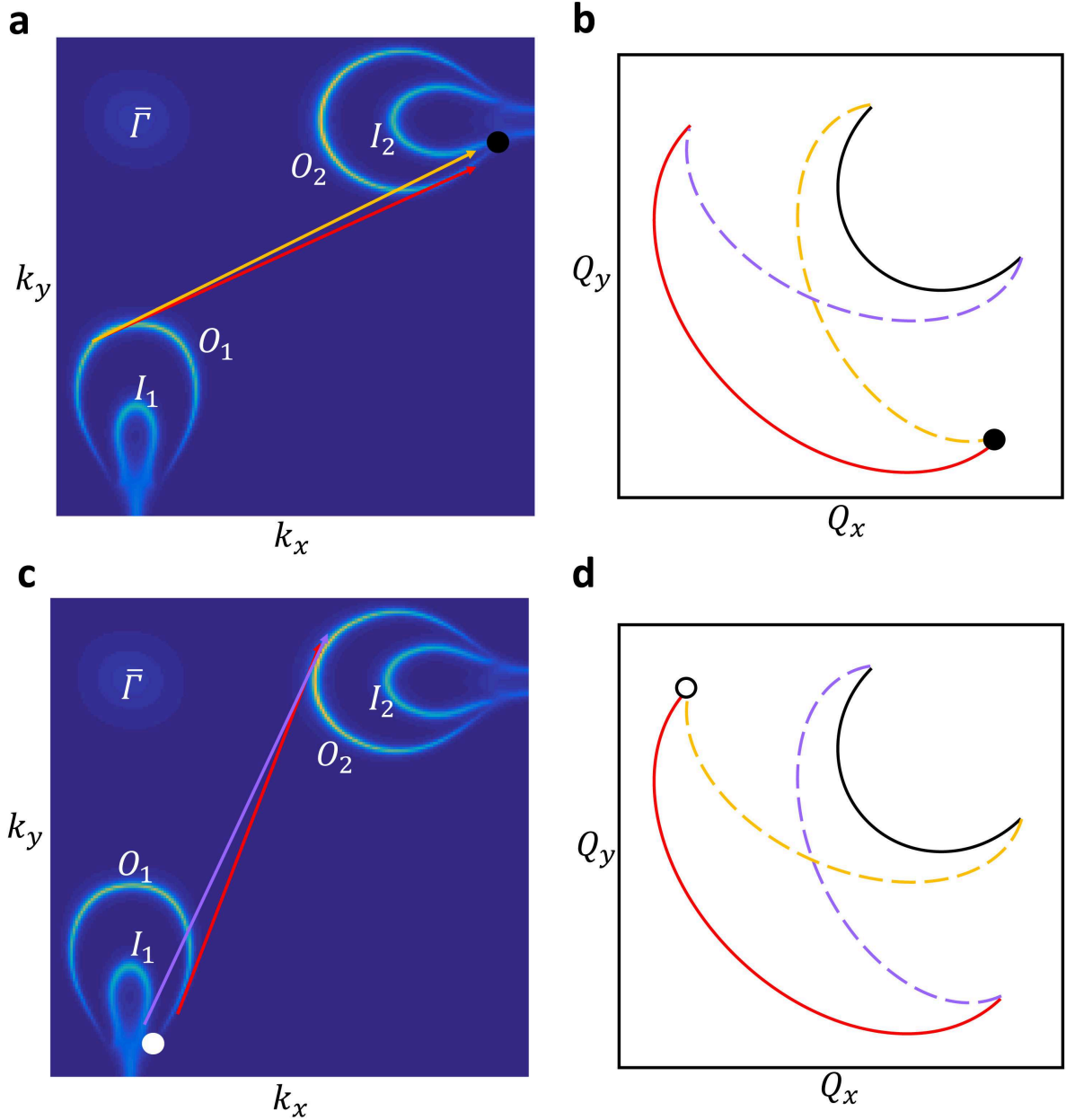}
\caption{\label{Nontrivial2}
\textbf{Fingerprints of the Weyl nodes in the interference pattern.}
(a) Calculated surface Fermi surface containing a pair of crescent Fermi arcs along $\bar{\Gamma}-\bar{X}$ and another pair along $\bar{\Gamma}-\bar{Y}$. The scattering vectors from outer(O1) arc to outer(O2) arc is noted by the red arrow. scattering vectors from outer(O1) arc to inner(I2) arc is noted by the orange arrow. The Weyl nodes near the $\bar{\Gamma}-\bar{X}$ axis is noted by a black dot.
(b) Schematic illustration of the QPI pattern based on the Fermi surface in panel (a). The red, orange, purple and black curves corresponds to the scattering between the outer(O1) and outer(O2) arcs, between outer(O1) and inner(I2) arcs, between inner(I1) and outer(O2) arcs, and between inner(I1) and inner(I2) arcs, respectively. (c,d) Same as panels (a,b). The only difference is that we consider the different scattering vectors.}
\end{figure*}

\begin{figure*}
\centering
\includegraphics[width=16cm]{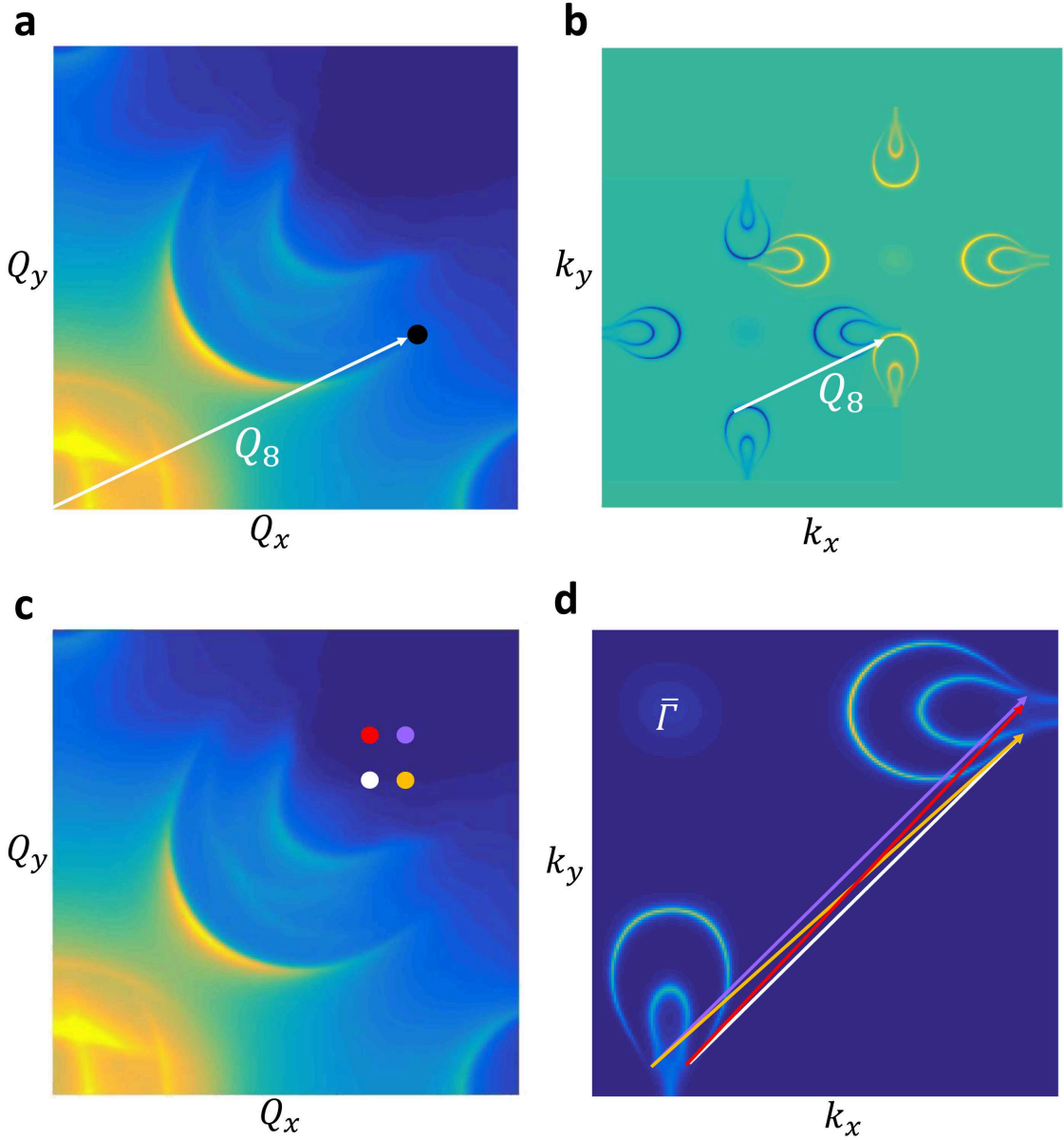}
\caption{\label{Nontrivial3}
\textbf{Scattering channel associated to the fingerprints of the Weyl nodes in the QPI pattern.}
(a) QPI feature that arise from the crescent Fermi arcs in one quadrant in $Q-$space. The scattering vector $Q_8$ corresponds to the termination-point in the QPI as noted by the black dot.
(b) Two copies of Fermi surfaces that are shifted by the vector $Q_8$ with respect to each other.
(c, d) The $Q$ points that correspond to the inter-Weyl-node scattering are shown by the dots in panel (c). The corresponding scattering vectors that connect the Weyl nodes are shown in panel (d).}
\end{figure*}

We now study the QPI features that arise from the crescent Fermi arcs. We note that on the top layer (As atoms), the crescent arcs along $\bar{\Gamma}-\bar{X}$ arise from $p_x$ orbital whereas those $\bar{\Gamma}-\bar{X}$ arise from $p_y$ orbital (Fig.~\ref{TaAsQPI}(b))). Hence if only the top layer were considered, the scattering between the crescent Fermi arcs would be suppressed. In other words, in order to observe the crescent Fermi arcs in the QPI, signals from the second (Ta) layer has to be significant in the STM data. As shown in Fig.~\ref{Nontrivial}(b), we find a complicated feature near the center of each quadrant in $Q$ space. This feature is due to the scattering between the crescent Fermi arcs along $\bar{\Gamma}-\bar{X}$ and those along $\bar{\Gamma}-\bar{Y}$, as noted by the scattering vectors $Q_4$ to $Q_7$. The zoomed-in view in Fig.~\ref{Nontrivial}(d) shows that it consists of four non-closed curves that join each other at four termination-points. We sketch a schematic for this feature in Fig.~\ref{Nontrivial2}(b). Among the four curves in Fig.~\ref{Nontrivial2}(b), the red curve is closest to the $Q$ space origin, meaning that it has the shortest $Q$ vector. Therefore, the red curve corresponds to the scattering between the two outer arcs as noted by the red arrow in Fig.~\ref{Nontrivial2}(a). Similarly, one can derive that the other three curves, namely, the black, orange, and purple curves come from the scatterings between the two inner arcs, between the outer (O1) and the inner (I2) arcs, and between the outer (O2) and the inner (I1) arcs, respectively. We now consider the meaning of end points. As shown in Fig.~\ref{Nontrivial2}(a), we start by considering the scattering from the outer(O1) arc to the outer (O2) arc noted by the red arrow; We move the ending point of the arrow through a Weyl node (the black dot) onto the the inner(I2) arc; Through this movement, the red arrow evolves into the orange arrow. Therefore, we see that the termination-point in Fig.~\ref{Nontrivial2}(b) is a fingerprint of the Weyl node in the QPI pattern because it corresponds to the scattering from a state on the outer(O1) arc to the Weyl node noted by the the black dot (Fig.~\ref{Nontrivial2}(a)). By the same token, it is straightforward to figure out that the other three termination points in Fig.~\ref{Nontrivial2}(b) are fingerprints of the other three Weyl nodes in Fig.~\ref{Nontrivial2}(a). Now let us again take the example of the black termination shown in Fig.~\ref{Nontrivial3}(a). As pointed out above, it corresponds to the scattering vector $Q_8$ which is from a state on the outer(O1) arc to the Weyl node. Then an obvious question is that which state on the outer(O1) arc is the starting point of this scattering. In order to understand that, we superimpose two copies of Fermi surfaces that are shifted in $k$ space by the vector $Q_8$. The overlapping areas between the two Fermi surfaces reveal the starting and ending points of the scattering. It can be seen from Fig.~\ref{Nontrivial3}(b) that the starting point of $Q_8$ is not a Weyl node but rather a $k$ point on the outer(O1) Fermi arc. To understand the starting state, we consider the following factors: When the two states connected by the $Q$ vector have the same direction of velocity, then the JDOS is maximized as the overlap between the two features is the largest. Indeed, this is roughly the case seen in Fig.~\ref{Nontrivial3}(b). Furthermore, the spin of the starting and ending states have to match. Therefore, the location of that starting state is determined by a complex consideration of both the joint density of states (JDOS) and the spin selection rule. In Figs.~\ref{Nontrivial3} (c,d), we further show the scattering vectors that directly connects two Weyl nodes, i.e., inter-Weyl-node scatterings. It can be seen that the termination-points do not coincide with the inter-node scattering.

\begin{figure*}
\centering
\includegraphics[width=16cm]{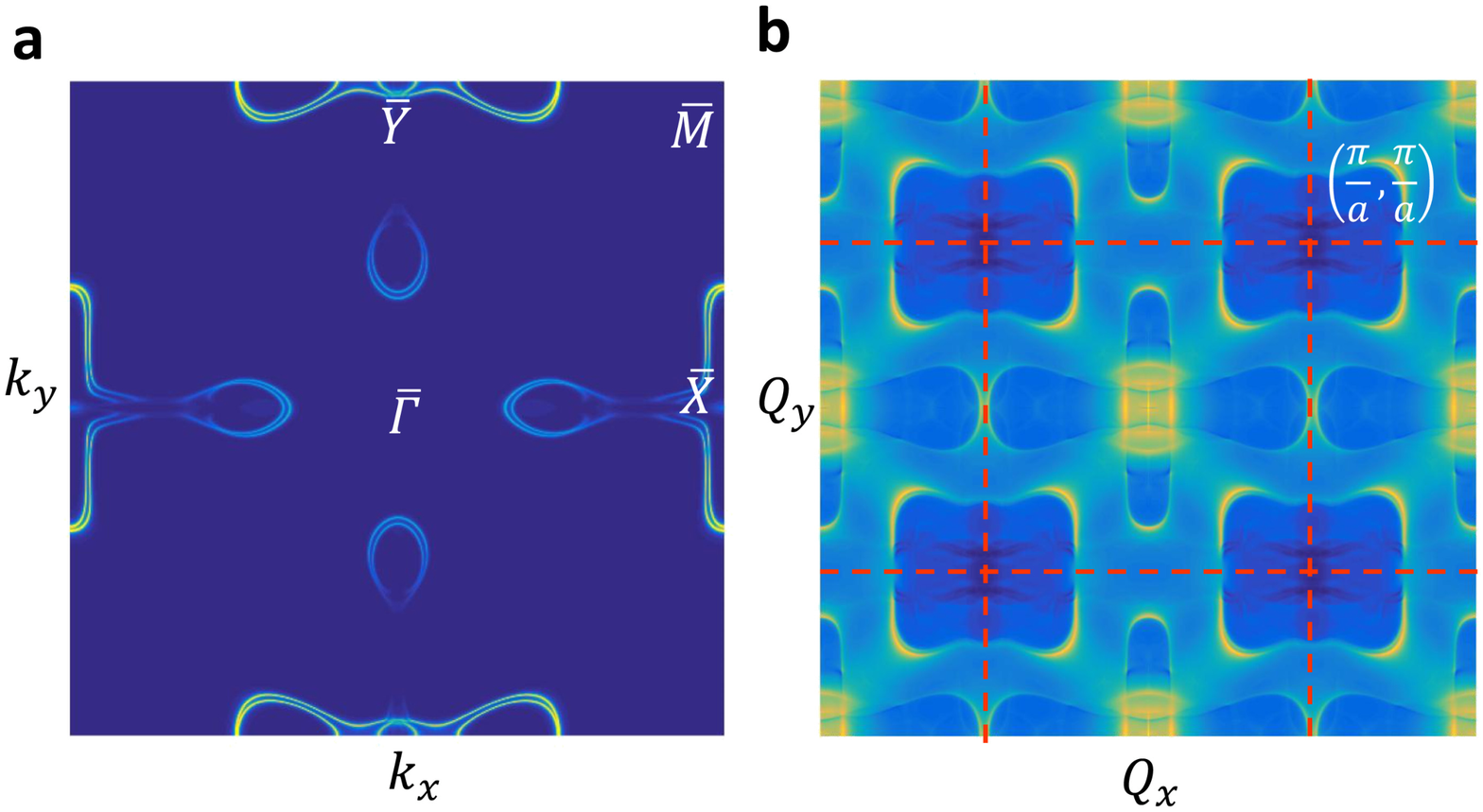}
\caption{\label{NbP}
\textbf{Quasi-particle interference pattern of the NbP(001) surface.}
(a) Calculated surface Fermi surface of NbP.
(b) Calculated QPI pattern of NbP based on the Fermi surface in panel (a).}
\end{figure*}

In Fig.~\ref{NbP}, we show the calculated Fermi surface and QPI of the NbP's surface states. It can be seen that the dominant features in the QPI pattern is qualitatively the same as TaAs. The only difference is that the features from the crescent Fermi arcs are too weak to be resolved in our calculation. This is because the spectral weight of the crescent Fermi arcs are quite low as seen in Fig.~\ref{NbP}(a). Systematic STM measurements of the QPIs of NbP are reported in Ref. \cite{Hao}. 

We compare the STM signature of Fermi arcs presented here with that of in angle-resolved photoemission spectroscopy (ARPES). In ARPES, one can show the existence of Fermi arcs without any theoretical calculations. Specifically, this can be done by counting the net number of chiral edgemodes along a closed $k$ loop that encloses a Weyl node in the surface electronic band structure in ARPES, as systematically discussed in Ref. \cite{ARPES1, TaP1, NbP1}. By contrast, what we found out here is that although the calculated QPI pattern shows fingerprints of the Fermi arcs and Weyl nodes, this is only achieved by referencing to the theoretical calculation. In other words, in an STM study, one will need to compare the STM data to the theoretical calculation and argue for the topological Fermi arcs based on the agreement between data and calculations. This is due to the indirect nature of STM measurements as it measures the momentum transfer rather than the real momentum. Although being indirect and less conclusive, the QPI patterns calculated here propose another evidence that is independent from the ARPES demonstrations \cite{ARPES1, TaP1, NbP1}, which is important for this rapidly developing field.

%Up to here, the scattering origin of the termination points in the QPI pattern has been understood. These termination points are fingerprints of the Weyl nodes in STM measurements.

%To this point, we have been following the logic that if we start from the calculated QPI, what are the features that are relevant to the Fermi arcs and the Weyl nodes. In other words, even if this were done by STM, one can only argue for a signature of the Fermi arcs \textbf{based on} the agreement with the theoretical calculations. Now, we would like to ask the question following the inverse logic sequence. That is, can we prove for the Fermi arcs and the Weyl semimetal state solely based on certain features in the STM (QPI) data? We note that this can be done in ARPES because one can count the net number of chiral edgemodes along a closed $k$ loop that encloses a Weyl node in the surface electronic band structure \cite{ARPES1, TaP1, NbP1}.

\begin{thebibliography}{99}

%\bibitem {rev1}
%M. Z. Hasan, S.-Y. Xu, and G. Bian, \textit{Phys. Scr.} \textbf{T164}, 014001 (2015).

\bibitem {rev2}
A. M. Turner and A. Vishwanath, http://arxiv.org/abs/1301.0330 (2013).


\bibitem{Weyl1} 
H. Weyl, \textit{I. Z. Phys.} $\mathbf{56}$, 330 (1929).

\bibitem{Weyl2}
L. Balents, \textit{Physics} \textbf{4}, 36 (2011).

\bibitem{Wan} 
X. Wan, A. M. Turner, A. Vishwanath, and S. Y. Savrasov, \textit{Phys. Rev. B} \textbf{83}, 205101 (2011).

\bibitem{Weyl3} 
A. A. Burkov and L. Balents, \textit{Phys. Rev. Lett.} \textbf{107}, 127205 (2011).

\bibitem{Weyl4} 
H. B. Nielsen and M. Ninomiya, \textit{Phys. Lett. B} \textbf{130} , 389 (1983)

\bibitem{Ojanen} 
T. Ojanen, \textit{Phys. Rev. B} \textbf{87}, 245112 (2013).

\bibitem{arc1} 
P. Hosur,  \textit{Phys. Rev. B} \textbf{86}, 195102 (2012).

\bibitem {arc2}
A. C. Potter, I. Kimchi, and A. Vishwanath, \textit{Nature Commun.} $\mathbf{5}$, 5161 (2014).


\bibitem{CA1}
C. Zhang, S.-Y. Xu, I. Belopolski, Z. Yuang, Z. Lin, B. Tong, G. Bian, N. Alidoust, C.-C. Lee, S.-M. Huang, T.-Rong, Chang, G. Chang, C.-H. Hsu, H.-T. Jeng, M. Neupane, D. S. Sanchez, H. Zheng, J. Wang, H. Lin, C. Zhang, H.-Z. Lu, S.-Q. Shen, T. Neupert, M. Z. Hasan, and S. Jia, http://arxiv.org/abs/1503.02630 (2015).

 
\bibitem{CA2} 
X. Huang, L. Zhao, Y. Long, P. Wang, D. Chen, Z. Yang, H. Liang, M. Xue, H. Weng, Z. Fang, X. Dai, and G. Chen, \textit{Phys. Rev. X} \textbf{5}, 031023 (2015)

\bibitem{Photon} 
C.-K. Chan, P. A. Lee, K. S. Burch, J. H. Han, and Y. Ran, http://arxiv:1509.05400 (2015).

\bibitem{TaAs1}
S.-M. Huang, S.-Y. Xu, I. Belopolski, C.-C. Lee, G. Chang, B. Wang, N. Alidoust,  G. Bian, M. Neupane, C. Zhang, S. Jia,  A. bansil. H. Lin, and M. Z. Hasan, \textit{Nature Commun.} \textbf{6}, 7373 (2015).

\bibitem {TaAs2}
H. Weng, C. Fang, Z. Fang, B. A. Bernevig, and X. Dai, \textit{Phys. Rev. X} \textbf{5}, 011029 (2015).

\bibitem{ARPES1}
S.-Y. Xu, I. Belopolski, N. Alidoust, M. Neupane, G. Bian, C. Zhang, R. Sankar, G. Chang, Z. Yuan,  C.-C. Lee, S.-M. Huang, H. Zheng,J. Ma, D. S. Sanchez, B. Wang, A. Bansil, F. Chou, P. P. Shibayev, H. Lin, S. Jia, and M. Z. Hasan, \textit{Science} \textbf{349}, 613 (2015).

\bibitem{ARPES2}
B. Q. Lv, H. M. Weng, B. B. Fu, X. P. Wang, H. Miao, J. Ma, P. Richard, X. C. Huang, L. X. Zhao, G. F. Chen, Z. Fang, X. Dai, T, Qian, and H. Ding, \textit{Phys. Rev. X} \textbf{5}, 031013 (2015).

\bibitem{ARPES3}
B. Q. Lv, N. Xu, H. M. Weng, J. Z. Ma,  P. Richard,  X. C. Huang, L. X. Zhao, G. F. Chen, C. E. Matt, F. Bisti, V. N. Strocov, J. Mesot,  Z. Fang, X. Dai, T. Qian, M. Shi, and H. Ding, {Nature Phys.} \textbf{11}, 724 (2015)

\bibitem{ARPES4}
L. X. Yang, Z. K. Liu, Y. Sun, H. Peng, H. F. Yang, T. Zhang, B. Zhou, Y. Zhang, Y. F. Guo, M. Rahn, D. Prbhakaran, Z. Hussain, S.-K. Mo, C. Felser, B. Yan, and Y. L. Chen, \textit{Nature Phys.} \textbf{11}, 728 (2015).


\bibitem{NbAs}
S.-Y. Xu, N. Alidoust, I. Belopolski, Z. Yuan, G. Bian, T.-R. Chang, H. Zheng, V. N. Strocov, D. S. Sanchez, G. Chang, C. Zhang, D. Mou, Y. Wu, L. Huang, C.-C. Lee, S.-M. Huang, B. Wang, A. Bansil, H.-T. Jeng, T. Neupert, A. Kaminski, H. Lin, S. Jia, and M. Z. Hasan, \textit{Nature Phys.} \textbf{11}, 748 (2015).

\bibitem{TaP1}
S.-Y. Xu, I. Belopolski, D. S. Sanchez, C. Guo, G. Chang, C. Zhang, G. Bian, Z. Yuan, H. Lu, Y. Feng, T.-R. Chang, P. P. Shibayev, M. L. Prokopovych, N. Alidoust, H. Zheng, C.-C. Lee, S.-M. Huang, R. Sankar, F. Chou, C.-H. Hsu, H.-T. Jeng, A. Bansil, T. Neupert, V. N. Strocov, H. Lin, S. Jia, M. Z. Hasan, \textit{Science Advances}, in press, http://arxiv.org/abs/1508.03102 (2015).

\bibitem{TaP2}
N. Xu, H. M. Weng, B. Q. Lv, C. Matt, J. Park, F. Bisti, V. N. Strocov, D. gawryluk, E. Pomjakushina, K. Conder, N. C. Plumb, M. Radovic, G. Aut\'es, O. V. Yazyev, Z. Fang, X. Dai, G. Aeppli, T. Qian, J. Mesot, H. Ding, and M. Shi, http://arxiv.org/abs/1507.03983 (2015).

\bibitem{TaP3}
Z. K. Liu, L. X. Yang, Y. Sun, T. Zhang, H. Peng, H. F. Yang, C. Chen, Y. Zhang, Y. F. Guo, D. Prabhakaran, M. Schmidt, Z. Hussain, S.-K. Mo, C. Felser, B. Yan, and Y. L. Chen, \textit{Nature Mater.} doi:10.1038/nmat4457 (2015).

\bibitem{NbP1}
I. Belopolski, S.-Y. Xu, D. Sanchez, G. Chang, C. Guo, M. Neupane, H. Zheng, C.-C. Lee, S.-M. Huang, G. Bian, N. Alidoust, T.-R. Chang, B. Wang, X. Zhang, A. Bansil, H.-T. Jeng, H. Lin, S. Jia, and M. Z. Hasan, http://arxiv.org/abs/1509.07465 (2015).

\bibitem{NbP2}
D. F. Xu, Y. P. Du, Z. Wang, Y. P. Li, X. H. Niu, Q. Yao, P. Dudin, Z.-A. Xu, X. G. Wan, and D. L. Feng, \textit{Chin. Phys. Lett.} \textbf{32}, 107101 (2015).

\bibitem{NbP3}
S. Souma, Z. Wang, H. Kotaka, T. Sato, K. Nakayama, Y. Tanaka, H. Kimizuka, T. Takahashi, K. Yamauchi, T. Oguchi, K. Segawa, Y. Ando http://arxiv.org/abs/1510.01503 (2015).
\bibitem{TaAs_spin1}
S.-Y. Xu, I. Belopolski, D. S. Sanchez, M. Neupane, G. Chang, K. Yaji, Z. Yuan, C. Zhang, K. Kuroda, G. Bian, C. Guo, H. Lu, T.-R. Chang, N. Alidoust, H. Zheng, C.-C. Lee, S.-M. Huang, C.-H. Hsu, H.-T. Jeng, A. Bansil, A. Alexandradinata, T. Neupert, T. Kondo, F. Komori, S. Shin, H. Lin, S. Jia, and M. Zahid Hasan, http://arxiv.org/abs/1510.08430 (2015).
\bibitem{TaAs_spin2}
B. Q. Lv, S. Muff, T. Qian, Z. D. Song, S. M. Nie, N. Xu, P. Richard, C. E. Matt, N. C. Plumb, L. X. Zhao, G. F. Chen, Z. Fang, X. Dai, J. H. Dil, J. Mesot, M. Shi, H. M. Weng, and H. Ding, http://arxiv.org/abs/1510.07256 (2015).


\bibitem{Hao}
H. Zheng, S.-Y. Xu, G. Bian, C. Guo, G. Chang, D. S.Sanchez, I. Belopolski, C.-C. Lee, S.-M. Huang, X. Zhang, R. Sankar, N. Alidoust, T.-R. Chang, F. Wu, T. Neupert, F.Chou, H.-T. Jeng, N. Yao, A. Bansil, S. Jia, H. Lin, and M. Z. Hasan, Atomic Scale Visualization of Quantum Interference on a Weyl Semimetal Surface. http://arxiv.org/abs/1511.02216 (2015).

\bibitem {HTC}
J. E. Hoffman, K. McElroy, D-H Lee, K.M. Lang, H. Eisaki, S. Uchida, and J.C. Davis, \textit{Science} \textbf{297}, 1148 (2002).

\bibitem {C}
R. M. Rutter, J.N. Crain, N.P. Guisinger, T. Li, P.N. First, and J. A. Stroscio, \textit{Science} \textbf{317}, 219 (2009).

\bibitem {STM1}
P. Roushan, J. Seo, C. V. Parker, Y. S. Hor, D. Hsieh, D. Qian, A. Richardella, M. Z. Hasan, R. J. Cava, and A. Yazdani, \textit{Nature} \textbf{460}, 1106 (2009).

\bibitem {STM2}
T. Zhang, P. Cheng, X. Chen, J.-F. Jia, X. Ma, K. He, L. Wang, H. Zhang, X. Dai, Z. Fang, X. Xie, and Q.-K. Xue, \textit{Phys. Rev. Lett.} \textbf{103}, 266803 (2009).

\bibitem {STM3}
I. Zeljkovic, Y. Okada, C.-Y. Huang, R. Sankar, D. Walkup, W. Zhou, M. Serbyn, F. Chou, W.-F. Tsai, H. Lin, A. Bansil, L. Fu, M. Z. Hasan, and V. Madhavan, \textit{Nature Phys.} $\mathbf{10}$, 572 (2014).





\end{thebibliography}
\end{document}